\documentclass[aps, prd, preprint, superscriptaddress, showpacs]{revtex4}
\usepackage{graphicx}

\usepackage{ulem}
\usepackage{epsfig}
\usepackage{color}
\definecolor{My_red}        {cmyk}{0.00, 1.00, 1.00, 0.20}

\newcommand{\bmat}{\left(\begin{array}}
\newcommand{\emat}{\end{array}\right)}
\newcommand{\beq}{\begin{equation}}
\newcommand{\eeq}{\end{equation}}
\newcommand{\wt}{\widetilde}

\def\ra{\rightarrow}

\def\Ld{\Lambda}
\def\ld{\lambda}
\def\f{\frac}

\def\bwt{\begin{widetext}}
\def\ewt{\end{widetext}}
\def\be{\begin{equation}}
\def\ee{\end{equation}}
\def\bea{\begin{eqnarray}}
\def\eea{\end{eqnarray}}
\def\bean{\begin{eqnarray*}}
\def\eean{\end{eqnarray*}}
\def\bary{\begin{array}}
\def\eary{\end{array}}
\def\bit{\begin{itemize}}
\def\eit{\end{itemize}}

\def\ra{\rightarrow}

\def\Ld{\Lambda}

\def\ld{\lambda}

\def\su5u1{SU(5) \times U(1)}
\def\fsu5u1{SU(5) \times U(1)'}
\def\so10{SO(10)}
\def\sq20{SO(10) \times SO(10)}

\def\ra{\rightarrow}

\def\Ld{\Lambda}
\def\ld{\lambda}
\def\f{\frac}

\def\L{\left(}
\def\R{\right)}
\def\bwt{\begin{widetext}}
\def\ewt{\end{widetext}}
\def\be{\begin{equation}}
\def\ee{\end{equation}}
\def\bea{\begin{eqnarray}}
\def\eea{\end{eqnarray}}
\def\bean{\begin{eqnarray*}}
\def\eean{\end{eqnarray*}}
\def\bary{\begin{array}}
\def\eary{\end{array}}
\def\bit{\begin{itemize}}
\def\eit{\end{itemize}}

\def\ra{\rightarrow}

\def\Ld{\Lambda}

\def\ld{\lambda}

\def\su5u1{SU(5) \times U(1)}
\def\fsu5u1{SU(5) \times U(1)'}
\def\so10{SO(10)}
\def\sq20{SO(10) \times SO(10)}

\usepackage[centertags]{amsmath}
\usepackage{amssymb}

\begin{document}

\title{Semi-direct Gauge-Yukawa Mediation }

\author{Zhaofeng Kang}

\affiliation{Key Laboratory of Frontiers in Theoretical Physics,
             Institute of Theoretical Physics,  Chinese Academy of Sciences,
             Beijing 100190,  P. R. China }

\author{Tianjun Li}

\affiliation{Key Laboratory of Frontiers in Theoretical Physics,
             Institute of Theoretical Physics,  Chinese Academy of Sciences,
             Beijing 100190,  P. R. China }

\affiliation{George P. and Cynthia W. Mitchell Institute for
             Fundamental Physics,  Texas A$\&$M University,
             College Station,  TX 77843,  USA }

\author{Tao Liu}

\author{Jin Min Yang}

\affiliation{Key Laboratory of Frontiers in Theoretical Physics,
             Institute of Theoretical Physics,  Chinese Academy of Sciences,
             Beijing 100190,  P. R. China }

\date{\today}

\begin{abstract}
We propose semi-direct  Gauge-Yukawa  mediation of
supersymmetry (SUSY) breaking. The messenger fields mediating
SUSY breaking to the visible sector do not directly couple
with the goldstino field,
and instead they have gauge and Yukawa interactions with some
primary messenger fields which couple directly with
the goldstino fields. From the explicit Feynman diagram
calculations for the SUSY breaking soft masses, we find that the SUSY particle
spectra can be realistic. Especially, this generalization of
semi-direct gauge mediation solves the massless gaugino mass
problem since the holomorphic soft mass terms of the messenger fields
can be generated by Yukawa couplings. We also provide some arguments that
this scenario can be realized naturally in some dynamical
SUSY-breaking models such as the ISS-like model.
\end{abstract}

\pacs{12.60.Jv,  14.70.Pw,  95.35.+d}

\maketitle

\section{Introduction and Motivation}

The mechanism for supersymmetry (SUSY) breaking is an open topic.
Due to the tree-level sum rule for the SUSY-breaking spectrum, namely
the supertrace theorem, SUSY must be broken (spontaneously) in the hidden
sector~\cite{Dimopoulos:1981zb}. Then how to mediate the SUSY-breaking
effects to the visible sector is a crucial issue. So far several schemes
have been proposed, among which the Gauge Mediated SUSY Breaking (GMSB)
is a quite promising approach because it can automatically avoid
the notorious flavor problem~\cite{early}
(for reviews, see, e.g., \cite{review}).
In particular, the direct
gauge mediation has recently attracted much attention, in which
the messengers play a role in determining the SUSY-breaking
vacua~\cite{Poppitz:1996fw}.
However, this attractive approach easily suffers from some problems,
for example, the  suppressed gaugino mass problem~\cite{Komargodski:2009jf},
the Landau pole problem, and the fine-tuning problem.
On the other hand, the indirect gauge mediation usually does not
have the gaugino mass problem but it has a demerit of adding
messengers by hand. A recent study on such messenger gauge mediation
is given in Ref.~\cite{Dumitrescu:2010ha}.

No matter direct or indirect gauge mediation, the messenger fields
$\Phi$ must carry the SUSY-breaking information in a form of the
mass splittings ($ m_{\phi, \pm}^2-M_{\phi}^2\neq 0$) between the
fermionic and bosonic components which are degenerate in a
SUSY-preserved theory. But how the messengers obtain such mass
splittings is unknown, and in the previous studies a singlet
$X=\langle X\rangle+\theta^2 F$ is simply introduced as the
SUSY-breaking source which couples directly with the messengers.
This is the so-called Minimal Gauge Mediation (MGM) and has been
widely considered in phenomenology.

\begin{figure}[htb]
\begin{center}
\includegraphics[width=6.0in]{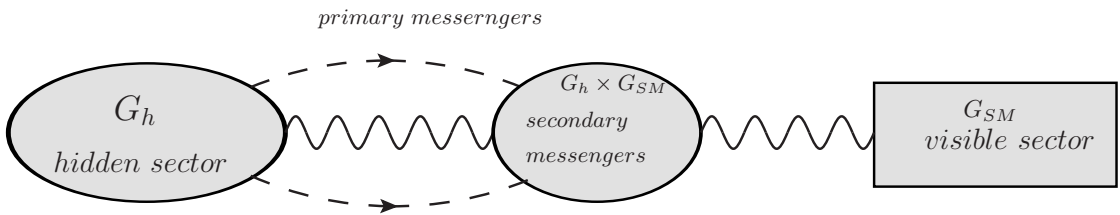}
\end{center}
\caption{\label{cascade} A schematic diagram showing the structure of the
two-step HCGYM.}
\end{figure}

Note that such a simplification in MGM may hide some questions,
for instance, why the messengers have to couple directly with $X$
to obtain mass splitting? In principle, the mass splitting can be
achieved in a cascade way, {\it i.e.}, some primary messengers
$\phi_0$ (neutral under the Standard Model (SM) gauge group
$G_{SM}$) mediate the SUSY-breaking to some secondary messengers
$\phi_1$ via the hidden sector gauge interaction with a gauge
group $G_h$ or Yukawa interaction, and this dynamical process can
be successive if necessary, until finally mediate the
SUSY-breaking to the visible sector. In this paper, we assume the
simplest cascade patten, {\it i.e.}, the two-step cascade
mediation, as depicted in Fig.~\ref{cascade}. In fact, such a
two-step cascade mediation via pure gauge interaction is nothing
but the semi-direct gauge mediation studied intensively in the
literature~\cite{Randall:1996zi,Seiberg:2008qj,moritz}. However, the
similar mediation mechanism with additional Yukawa interaction has
not been studied, which will be examined in this article.

The conventional  semi-direct gauge mediation suffers massless gaugino mass
problem, since the secondary messengers does not acquire holomorphic
soft terms~\cite{Ibe:2009bh,Argurio:2009ge}. However, with
Yukwa-assisted mediation, the holomorphic soft terms
arise at one loop, thus gauginos can indeed become massive.
We shall show that a calculable cascade SUSY-breaking with the
minimal K\"ahler potential, based on a very general setup,
is viable only when both gauge and Yukawa
interactions are  turned on. Namely, this hybrid cascade gauge-Yukawa
mediation (HCGYM) can provide the viable SUSY particle spectra in
some cases. The crucial point is the presence of non-vanishing
supertrace in the messenger sector which links to the visible sector
and modifies the sfermion masses drastically.

Note that our cascade scenario has one essential difference from the
 gauge mediation with messenger sector  added by hand, {\it i.e.}, our
scenario can be naturally realized dynamically such as in the
Intriligator-Seiberg-Shih (ISS)-like models~\cite{Intriligator:2006dd}.
In other words, the messengers in our scenario are
a natural part in the hidden sector which breaks SUSY. We simply
take a different way to gauge the so-called global flavor symmetry
in the hidden sector. However, in our study we will start from a
class of weakly coupled effective O'Raifeartaigh (OR) models,
characterized by cubic terms and some large global symmetry $\supset
G_h\times G_{SM}$ (some will be gauged properly), without any
elaboration on their dynamical realization.

The paper is organized as follows. In Section II we propose a
general framework for HCGYM and then calculate the
SUSY breaking soft terms via the secondary messengers.
Our calculations will be performed
explicitly in a very general setup. In Section III we analyze
in what conditions the realistic soft masses in the visible sector can be
achieved, and comment on the possible general features of the HCGYM.
Finally,  discussions and conclusion are given in Section
IV.

\section{Hybrid Cascade Gauge-Yukawa Mediation}

\subsection{The Primary SUSY-Breaking Mediation}

First we focus on SUSY breaking mediation without specifying any concrete model.
Generally, a cascade mediation via gauge and Yukawa
 (hyprid) interactions has a typical structure as follows
\begin{itemize}
\item[(i)] A hidden sector, which breaks SUSY,  is composed of Goldstino
superfield $X$ and the primary vector-like messenger fields
$(\phi_i, \bar \phi^i)$ with $i$ being the gauge/flavor index of the
hidden sector group  $G_h$ like $SU(N)$.
\item[(ii)] The secondary messenger fields $(F, \bar F)$, which are charged under
$G_h$ and the SM gauge group, and the vector-like fields $(f, \bar f)$,
which only carry the SM quantum numbers, both have supersymmetic mass terms
and obtain SUSY-breaking effects via the interaction with the hidden gauge
fields as well as from the  cubic Yukawa coupling terms (discussed below).
\item[(iii)] Some cubic terms must be present, which couple one
primary and one secondary messenger fields with the SM vector-like fields,
or couple the two secondary
messengers with one primary messenger. However, the cubic terms with two
primary fields are not allowed by the SM gauge invariance.
Note that if we turn off these cubic terms, the hidden sector SUSY-breaking
will not be affected due to the fact that the secondary messengers are not
relevant to the SUSY-breaking dynamics although they couple with the hidden
sector. Then our scenario will just be reduced to the
semi-direct gauge mediation discussed in Ref.~\cite{Seiberg:2008qj}.
\end{itemize}
We stress that our scenario is significantly different from the
semi-direct gauge mediation~\cite{Seiberg:2008qj}. The semi-direct
gauge mediation \cite{Seiberg:2008qj} interpolates the MGM and the
direct gauge mediation. It introduces a messenger sector that does
not affect the SUSY-breaking and does not have to couple with $X$.
However, it uses the messenger sector to probe the hidden sector
only via some hidden gauge interactions. In our approach,  the
direct renormalizable couplings in the superpotential play a
central role. As a result, the spectrum of the secondary
messengers and also the visible fields is modified greatly. The
presence of such cubic terms can arise in  dynamical models, and,
moreover, the mass scale can be dynamically determined. In other
words, our framework regards the messenger sector as a built-in
part of the dynamical model building, rather than introduced by
hand.


With the above general setup, we propose to use the effective OR model which
shows the dynamical structure. In our convention, we choose a basis in which
the light messengers are diagonal while the hidden sector takes a general form
\begin{align}
W=&\left[{\cal
F}X+(\lambda_{ab}X+m_{ab})\bar\phi_{a}^i\phi_{bi}\right] \cr
&+\L\lambda_{a}\phi_{ai}\bar F^if+\bar{\lambda}_{a}\bar\phi_{a}^i
F_i\bar f\R+\L M_FF_i\bar F^i+M_ff\bar f\R.
\end{align}
Here, the index $a$ does not refer to any symmetry and we require
$M_{F, f}\ll max\{\langle X\rangle, m\}\sim M_h$, which is defined as
the typical mass scale in the hidden sector, while $\sqrt{{\cal F}}$
determines the SUSY-breaking scale.  Note that the multiple pairs of
secondary messengers have additive contributions and thus do not
change our conclusion. By the way, we may embed the SM gauge group
into the group $G_h$ and then it induces the massless gaugino mass problem
\cite{Komargodski:2009jf}
\begin{align}
M_{\ld}\propto \f{\partial }{\partial X}\log {\rm det}{\cal
M}^{primary}_F=0,
\end{align}
where ${\cal M}^{primary}_F$ denotes the supersymmetric mass matrix for
the hidden sector messenger.

Note that our general setup can be equipped with the $R$-symmetry.
With the $R$-symmetry our framework can be considered as a
generalization of the (extra)ordinary GMSB~\cite{Cheung:2007es}
with the cubic terms and the extra gauge group $G_h$. Then the
particle charges under $R-$symmetry are
\begin{align}
&R(X)=2,\quad R(F)+R(\bar F)=2,\quad R(f)+R(\bar f)=2, \cr
 &R(\bar \phi_a)+R(\phi_b)=0\,\, ({\rm for\,\,}\ld_{ab}\neq0),
 \quad R(\bar \phi_a)+R(\phi_b)=2\,\, ({\rm for\,\,}m_{ab}\neq0),\cr
 &R( \phi_a)+R(\bar F)+R(f)=2\,\, ({\rm for\,\,}\ld_{a}\neq0),
 \quad R( \bar\phi_a)+R( F)+R(\bar f)=2\,\, ({\rm
 for\,\,}\bar{\ld}_{a}\neq0).
\end{align}
To make gauginos massive, $R-$symmetry must be broken.
Here, we will not scrutinize the $R-$symmetry breaking mechanism.
Instead, we simply assume it can be realized by $\langle X\rangle\neq 0$.
This can be achieved by introducing some proper structures
in the hidden sector, leading to the spontaneous breaking
of the $R-$symmetry radiatively~\cite{Shih:2007av} or at
the tree-level~\cite{Sun:2008va,Komargodski:2009jf}.

Next, we discuss the SUSY-breaking encoded in the secondary messengers.
They feel the SUSY-breaking effects through the following ways.
One is through the conventional gauge mediation at two-loop level, of the
non-holomorphic form such as  $m_{F.g}^2|F|^2$ (Because $F$ and $f$ are
similar, for simplicity, we will only focus on the former.).
They also receive the direct one-loop contributions from the
Yukawa couplings, which are calculated explicitly in Appendix ~\ref{2}.
Moreover, there are  two-loop contributions, which
are $\propto \ld_a^4$ and can be extracted using the wave function
renormalization method~\cite{Giudice:1997ni,holo}.
In short, these three terms are given by
\begin{align}\label{sfg}
m_{F, g}^2\simeq&\frac{2g_h^4}{(16\pi^2)^2}\frac{{\cal
F}^2}{M_h^2}\frac{N^2-1}{2N},\\
m^2_{F,Y1}=&m^2_{ F F^*}=i(\Sigma_4+\Sigma_5+\Sigma_6), \\
m^2_{F,Y2}=&m^2_{FF^*}=\frac{\lambda_a^4}{(16\pi^2)^2}(N+2)\frac{F^2}{M_h^2},
\end{align}
where $g_h$ is the gauge coupling of $G_h$ and its value will be constrained
by phenomenology. The explicit expressions of $\Sigma_i$ functions are given
in Eqs.~(\ref{EqnA}), (\ref{EqnB}), and (\ref{EqnC}).
Such terms enter into the diagonal elements of the
bosonic components of the secondary messengers, and behave as
the $D-$term SUSY-breaking.
In additional to the non-holomorphic terms, the other soft terms
are holomorphic terms, namely $m_{F\bar F}^2F\bar F+h.c.$, which is
generated at one-loop level due to the Yukawa interaction and proportional
to the secondary messenger masses
\begin{align}
m_{hol}^2\equiv m^2_{F\bar F}=&m^2_{F^*\bar
F^*}=i(\Sigma_1+\Sigma_2+\Sigma_3)\propto M_F~,~
\end{align}
where we have set $M_F\simeq m_f$.  This term contributes to the
non-diagonal mass terms of the secondary messengers and acts as the
conventional $F-$term SUSY-breaking. It is absent in conventional
semi-direct gauge mediation, rendering gaugino massless at the leading
order of SUSY-breaking.

A comment is in order. For the SDGM there is a robust  gaugino
screen theorem \cite{holo}. It is derived through the renormalization
of the real physical superfield
\begin{align}
R(\mu)=S(\mu)+S^\dagger(\mu)+\f{T_G}{8\pi^2}\log {\rm Re(S(\mu))}
-\sum_r\f{T_r}{8\pi^2}\log {\cal Z}_r(\mu),
\end{align}
where $S(\mu)$ is the holomorphic gauge coupling that runs only at
one-loop and $r$ runs over light fields at scale $\mu$. The
screening  depends heavily on the replacement
$\mu_{F}^2=\frac{M_FM_F^\dagger}{\mathcal {Z}_F^2(\mu_F)}$ after
across all messenger scales, leading to $R(\mu)$ independent on $X$
at leading order. However,  just like the analysis in \cite{Argurio:2010fn}
which introduced chiral messengers so that the two chiralities 
of the messengers have different wave functions and thus invalidate 
the replacement, in our case the chiral Yukawa coupling between the
secondary and primary messengers can also invalidate the replacement.
Consequently, the screening is avoided.

By the way, in our framework we assume that $G_h$ is not
Higgsed. If it is Higgsed, the gauge mediation has some
modification and the modification extent is controlled by
a new parameter $y=M_{V}^2/M_h^2$ with $M_V$ being the vector boson
mass scale~\cite{Gorbatov:2008qa}. When it tends to be zero, we have the
complete gauge group, while, oppositely, we have no gauge symmetry at
the messenger scale. A sufficiently high $M_V$ is not a desired case
and the reason will be discussed in the following.

\subsection{Gaugino and Sfermion Masses}

In the above we have studied the SUSY-breaking mediated to the
secondary messengers by hybrid mediation. Now we calculate the
SUSY-breaking mediated to the visible sector by the conventional
GMSB. In contrast to the simple GMSB model where the hidden sector
messenger spectrum has a vanishing supertrace, in our cascade
framework the supertrace is nonzero due to the radiative
corrections which are essentially from the holomorphic soft mass
term  $m^2_{F\bar F}$. In the basis $(F,\bar F)$, the mass matrix
of the scalar component is
\begin{align}
{\cal M}_B^2=\left(\begin{array}{cc}
  M_F^2+D & F \\
  F & M_F^2+D
\end{array}\right),
\end{align}
where $F=m_{hol}^2$  can be  effectively treated as a spurion
superfield $X$ which has a vanishing
vacuum expectation value (VEV) for the lowest component but
has a non-zero $\theta^2$ component. From this point of view, our
framework provides a natural realization of multi-spurion fields,
which will be helpful to implement general gauge
mediation~\cite{Meade:2008wd}. The term $D=m_{nhol}^2$ measures the
supertrace. It has deep implication in the soft masses in the
visible sector. The explicit calculations have been carried out
in Ref.~\cite{Poppitz:1996xw}, and we use their results to
calculate the soft
masses of gauginos and sfermions.

In the limit of small SUSY-breaking $D,F\ll M_F^2$,
the generic sfermion mass sqaure is approximately given by
\begin{align}\label{D}
m_{\wt f}^2\simeq&
\sum_aC_a\L\frac{\alpha_a}{4\pi}\R^2\left[2\frac{F^2}{M_F^2}+D\L
-\frac{2}{3}\frac{F^2}{M_F^4}
-4\log\frac{\Lambda^2}{M_F^2}+4\R\right],
\end{align}
where $C_a$ is the Dykin index of $\wt f$ related to the three SM
gauge groups. $\Lambda$ is the ultraviolet cutoff scale related to a
high scale at which $D$ and $F$ are generated, which, in fact, is
the primary messenger scale $\Lambda \sim M_h\gg M_{F,f}$. This
logarithmic UV-dependence is a character of the cascade gauge
mediation. As for the gaugino masses, the $D-$type contribution does
not affect their masses significantly, and thus they take a
conventional form
 \begin{align}\label{DG}
M_{\ld_a}\simeq& \frac{\alpha_a}{4\pi}\frac{F}{M_F}~.~\,
\end{align}
Now some comments are due regarding the property of the soft
masses in the presence of $D-$type contribution.
>From Eq.~(\ref{D}) we obtain
\begin{align}\label{DA}
m_{\wt f}^2\simeq&
\sum_aC_a\L\frac{\alpha_a}{4\pi}\R^2\left(2\frac{F^2}{M_F^2}-4D
\log\frac{\Lambda^2}{M_F^2}\right).
\end{align}
So, depending on the sign of $D$, the sfermion mass square can be
either enhanced ($D<0$) or reduced ($D>0$). As calculated in the
previous section,  there are several sources of non-holomorphic
terms, and in the following we will show that a pure Yukawa
cascade model is not viable practically.

First, in the cascade gauge mediation, a contribution from pure
gauge interaction is necessary.  In the limit $g_h\ra 0$,
generically the one-loop effect is dominant and scales as (the index
$a$ will be dropped for simplicity)
\begin{align}
 D=m_{FF^*}^{(1)}&\sim
{\rm sign(D)}\times\frac{\lambda^2}{16\pi^2}\frac{{\cal
F}^2}{M_h^2},
\end{align}
we will find that the sign is not predicted uniquely  but dependent
on the parameters in the hidden sector. On the other hand, the
holomorphic term sales as
\begin{align}
 F&\sim
\frac{\lambda^2}{16\pi^2}\frac{{\cal F}}{M_h}\times M_F,
\end{align}
which is suppressed by the light supersymmetic mass term of the
secondary messengers. Then we find that the $D-$type contribution
dominates over the sfermion mass term. Concretely, in Eq.~(\ref{D})
the second term is about
$\L\log\frac{\Lambda}{M}\R{64\pi^2}/\lambda^2$ times larger than the
first term which is at the same order of gaugino mass.  As a result,
it leads to a splitting spectrum, which will incur large fine-tuning
in the Minimal Supersymmetric Standard Model (MSSM).

The above conclusion is obtained only when the holomorphic and
non-holomorphic SUSY-breaking terms are generated at the same loop
level. We note that in some cases the two-loop contributions can
dominate over the one-loop contributions which is suppressed by
higher order of SUSY-breaking $\propto {{\cal F}^4}/{M^6_h}$ due to
some subtle cancellation~\cite{Poppitz:1996xw}. This happens when
the hidden sector essentially contains only one mass scale $\propto
{{\cal F}}/{M_h}$ which is irrelevant to the Yukawa couplings between
$\Phi$ and $X$, as in the minimal GMSB with $X$ being the spurion
field. In our model, this is approximately satisfied in the limit
$\langle X\rangle\gg m$, but, unfortunately, a realistic spectrum cannot
be obtained. In fact, the $D-$type contribution scales as
\begin{align}
D\sim \frac{1}{(16\pi^2)^2}\lambda^4\frac{{\cal F}^2}{M_h^2}.
\end{align}
Then the sfermion mass is given by
\begin{align}
m_{sfermion}^2\sim
\L\frac{\alpha_a}{4\pi}\R^2\frac{\lambda^4}{16\pi^2}\frac{{\cal
F}^2}{M_h^2} \left[2-4(N+2)\log\L\frac{\Lambda^2}{M^2_h}\R\right]
\end{align}
which is always negative and unacceptable since it breaks
the  SM gauge group $G_{SM}$ at the messenger scale.

Now we consider the second limit $\ld,\bar {\ld}\ra 0$ with a
nonzero $g_h$. In this case we recover the conventional semi-direct
gauge mediation scenario.  The study in Ref.~\cite{Ibe:2009bh,Argurio:2009ge}
pointed out that, independent of the
details of the hidden sector, the doubly charged messengers do not
acquire $F-$type SUSY-breaking information from it and thus the
gaugino mass vanishes at the leading order (gaugino screening).
In Ref.~\cite{Argurio:2010fn} some complicated chiral messener
fields are proposed
to overcome this difficulty. In our HCGYM approach we try to solve
this by turning on the Yukawa couplings.

In our HCGYM scenario, we find that for a proper value of $g_h/\ld$,
the sfermion mass square can be reduced as a result of non-vanishing
supertrace. The sfermion masses receive a $D-$type contribution as
\begin{align}
 {\rm sign}(D)\frac{\lambda^2}{16\pi^2}\frac{{\cal
F}^2}{M_h^2}+\frac{2g_h^4}{(16\pi^2)^2}\frac{{\cal
F}^2}{M_h^2}\frac{N^2-1}{2N}.
\end{align}
Therefore, for ${\rm sign}(D)=-1$, there exists cancellation between
the gauge and
Yukawa contributions. This $D-$type contributions approximately vanishes
and the sparticle spectrum reduces to the MGM case when
\begin{align}
\frac{g_h}{4\pi}\sim\L\frac{\lambda}{4\pi}\R^{1/2}\L{\frac{N}{N^2-1}}\R^{1/4}.
\end{align}
In fact, we can tune the couplings to obtain our required soft masses.
Note that the reduced sfermion spectrum generically requires a
tuning of the ratio $\ld/g_h$, which arises from the cancellation
between the one-loop effect ($\propto \ld^2$) and the two-loop
effect ($\propto g_h^4$). By the way, increasing $N$ helps to lower
the value of $g_h$, which, however, may induce the Landau pole for
$SU(3)_C$ gauge coupling.

If the two-loop Yukawa corrections are dominant, different results will be obtained.
Then the $D-$type contribution  is
\begin{align}
\frac{1}{(16\pi^2)^2}\lambda^4\frac{F^2}{M_h^2}+\frac{2g_h^4}{(16\pi^2)^2}\frac{{\cal
F}^2}{M_h^2}\frac{N^2-1}{2N}.
\end{align}
Due to the fact that both $D-$ and $F-$type contributions come
from the two-loop level, they have the same sign and the additional
gauge contribution  makes the sfermion mass quite large (negative).

Our above analysis is just a rough estimation. In the next subsection
we will give a concrete example and perform the numerical calculations to
show that our framework indeed works.

\subsection{A Concrete Example and A Preliminary Dynamical Realization}

\begin{figure}[htb]
\begin{center}
\includegraphics[width=3.5in]{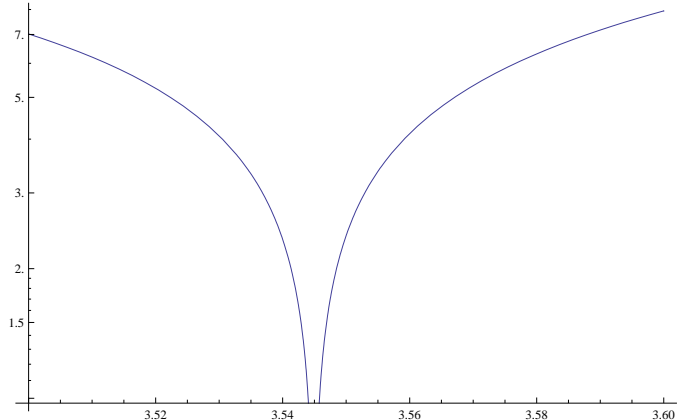}
\end{center}
\vspace*{-0.5cm} \caption{\label{ratio}. The ratio of $D-$type and
$F-$type contributions $|R|$,  as a function of $g_h$ . Here we take
$N=4$ and $\ld=1$, $\Ld/M_F=10^4$. The right-handed half curve
covers the region with  $R<0$.}
\end{figure}

To show that our HCGYM scenario can satisfy the required property, we now
study numerically a concrete example. For this simple example the
superpotential is assumed to be
\begin{align}
W=X\bar\phi_{1}\phi_{1}+\mu(\bar\phi_1\phi_2+\bar\phi_2\phi_1)+\lambda\phi_2F\bar
f+\lambda\bar\phi_2\bar Ff+m_FF\bar F+m_ff\bar f,
\end{align}
where $X$ is regarded as the spurion superfield, both $(\phi,\bar \phi)$ and
$(F,\bar F)$  belong to the fundamental representation of the
hidden gauge group $SU(N)$ while $(f,\bar f)$ is neutral. We set
$\langle X\rangle=M+F_X\theta^2$ with $M=1$, $\mu=1$ and $F_X=0.1$.
Then we get the $D$ and $F$ terms as
\begin{align}
D=m_{FF^*}=&-0.003\times\frac{1}{16\pi^2}\lambda^2+
0.008\times\frac{g_h^4}{(16\pi^2)^2}\frac{N^2-1}{2N},\\
F=m_{F\bar F}^2=&0.07\times\frac{1}{16\pi^2}\lambda^2M_F.
\end{align}
Another messenger pair $(f,\bar f)$ also obtain soft $D$- and $F$-type
corrections which are enhanced by a group factor $N$, but receive
no corresponding gauge contribution since they are charged under
$G_{SM}$ only. With all these contributions, from Eq.~(\ref{D}) we
find that for the sfermion masses the $D-$type contribution and
$F-$type contribution has a ratio:
\begin{align}
R=-2(1+N)\log\frac{\Lambda^2}{M_f^2}\times\frac{(-16\pi^2\lambda^2)
+g_h^4\frac{N-1}{2N}0.008}{(0.07\lambda^2)^2(N^2+1)}.
\end{align}
Note that this result is independent of $M_F$. In Fig.~\ref{ratio}
we plot this ratio as a function of $g_h$ (setting $\ld=1$). As
pointed previously, the ratio can be negative and thus can reduce
the sfermion masses with some tuning of $\ld/g_h$.

Let us comment on the dynamical realization of our framework.
In principle, any conventional semi-direct gauge mediation
models can be updated to our HCGYM scenario by coupling the secondary
messengers to the hidden sector. Another more natural model can be
found in Ref.~\cite{Amariti:2010sz}, which studied the metastable
SUSY-breaking vacua in the $SU(N_c)$ supersymmetric QCD (SQCD) (${\cal N}=1$) in the
superconformal window $3/2N_f<N_c<2N_f$ with $N_f$ being the number
of the vector-like quarks. Based on the SQCD in Ref.~\cite{Barnes:2004jj},
with some deformations, Ref.~\cite{Amariti:2010sz} studied its Seiberg dual description
\begin{align}
W= Kp\wt q+ L\wt pq+ Nq\wt q+M_p p\wt p-\mu^2N.
\end{align}
The flavor symmetry is $SU(N_f^{(1)})\times SU(N_f^{(2)})$ and the
magnetic gauge group is $SU(\wt N)$ with $\wt N\equiv N_f^{(1)}+N_f^{(2)}-N_c$.
Various fields are assigned in the representation as
\begin{align}
\begin{array}{c||c|c|c}
              & SU(N_f^{(1)}) & SU(N_f^{(2)}) & SU(\wt N) \\\hline
             N & \bar {N_f}^{(1)}\otimes N_f^{(1)} & 1 & 1 \\\hline
             q+\wt q & \bar {N_f}^{(1)}\oplus N_f^{(1)}& 1 & \wt N\oplus\bar{\wt N}
             \\
             p+\wt p & 1 & \bar{N_f}^{(2)}\oplus N_f^{(2)} & \wt N\oplus\bar{\wt
             N}\\\hline\hline
                           K+L & \bar {N_f}^{(1)}\oplus N_f^{(1)} & \bar {N_f}^{(2)}\oplus N_f^{(2)} &
             1
           \end{array}~.~
\end{align}
Aside from the last line, this just gives the content of the
ISS model~\cite{Intriligator:2006dd}. It has a SUSY-breaking
vacuum, with superfields parameterized as
\begin{eqnarray}
&& q=\left(\begin{array}{c}
\mu+\sigma_1 \\
\phi_1\end{array} \right), \quad \tilde q = ( \begin{array}{cc}
\mu+\sigma_2&\phi_2 \end{array}),\quad N=\left( \begin{array}{cc}
\sigma_3&\phi_3\\
\phi_4 &X
\end{array}\right),\nonumber \\
&& p=\phi_5,\quad \tilde p = \phi_6,\quad
L=(\begin{array}{cc}\phi_7&\tilde Y\end{array}), \quad K=\left(
\begin{array}{c}
\phi_8\\ Y \end{array} \right).
\end{eqnarray}
Expanding around this vacuum, we obtain
\begin{align}\label{mag}
W= &\L
X\phi_1\phi_2-\mu^2X+\mu(\phi_1\phi_4+\phi_2\phi_3)\R+\mu(\phi_5\phi_8+\phi_6\phi_7)\cr
&+\L Y\phi_2\phi_5+\wt Y\phi_1\phi_6\R+M_p\phi_5\phi_6.
\end{align}
As shown in Ref.~\cite{Amariti:2010sz}, when the flavor symmetry satisfies the relation
\begin{align}
N_f^{(2)}<2\wt N<N_f^{(1)}+N_f^{(2)}<3\wt N,
\end{align}
the theory is weakly coupled at IR and has tree-level
SUSY-breaking vacua. Choosing $N_f^{(2)}=5,N_f^{(1)}=2,\wt N=3$,
and embedding $G_{SM} \subset SU(N_f^{(2)})$, we find that
$\phi_{1-4}$ play the role of primary messengers while
$\phi_{5-8}$ are secondary messengers. From Eq.~(\ref{mag}) it can
be clearly seen that the hidden sector SUSY-breaking effect
mediated by gauge and Yukawa interactions to the secondary
messengers indeed appear in the model. Of course, it is not a
realistic model since the $R-$symmetry is not broken. To make it
realistic, further modifications in the hidden sector are
required, which is beyond the scope of this paper.

\section{Discussion and Conclusion}

We proposed a pattern of cascade gauge-Yukawa mediation of SUSY-breaking
based on an explicit Feynman diagram calculations for the soft mass terms.
We found that the realistic soft masses for gauginos and sfermions
 can be obtained in the case that
the conventional messenger fields obtain the SUSY-breaking effects through
gauge and Yukawa interactions with the hidden sector. This hybrid scenario
can easily avoid the massless gaugino mass problem and reduce the sfermion
masses by some fine-tuning between $\ld$ and $g_h$.

Finally, we note that very recently such a cascade patten was
utilized for SUSY breaking (not mediation) in Refs.~\cite{Ibe:2010jb,McCullough:2010wf}.
And the $R-$symmetry spontaneous breaking at the two-loop
level~\cite{Giveon:2008ne,Amariti:2008uz} is also a result of
cascade effect.

\section*{Acknowledgments}
This work was supported by the National Natural Science Foundation of
China under grant Nos. 10821504, 10725526 and 10635030,
by the DOE grant DE-FG03-95-Er-40917, and by the
Mitchell-Heep Chair in High Energy Physics.

\section{Appendix}
\subsection{Soft Mass Terms from the One-Loop Yukawa Mediation }
\label{2}

Here we present the explicit one-loop calculations for
the soft terms of the secondary  messengers, using dimensional
regularization. First of all, the mass square matrix of the primary
messenger bosonic components ($\phi, \bar\phi^*$) and the  Dirac
fermion mass matrix are respectively given by
\begin{align}
{\cal M}^2_{B}&=\left(\begin{array}{cc} (\lambda
X+m)^\dagger(\lambda X+m) & \lambda {\cal F}
\\ (\lambda  {\cal F})^\dagger & (\lambda X+m)^\dagger(\lambda
X+m)\end{array}\right),\\ {\cal M}_{F}&=\lambda X+m,
\end{align}
where ${\cal M}_F$ is a $N_p\times N_p$ matrix while ${\cal M}^2_{B}$
is $2N_p\times 2N_p$. They can be diagnolized by the unitary
matrix $U$, $ N_L$ and $ N_R $ respectively
\begin{align}
U^+{\cal M}^2_{B}U=M_{1},\\
 N_L{\cal M}_FN_R=M_{2}.
\end{align}
For the convenience of our following calculations, we parameterize
the matrix in a block form
\begin{align}
U=\left(\begin{array}{cc}A&B\\C&D\end{array} \right),\quad
M_1=\left(\begin{array}{cc}m_1\\m_2\end{array} \right).
\end{align}
Then the soft terms are given by
\begin{align}
V_{soft}=(m^2_{F\bar F}F\bar F+h.c.)+m^2_{FF^*}FF^*+m^2_{\bar F\bar
F^*}\bar F\bar F^*,
\end{align}
\begin{figure}[htb]
\epsfig{file=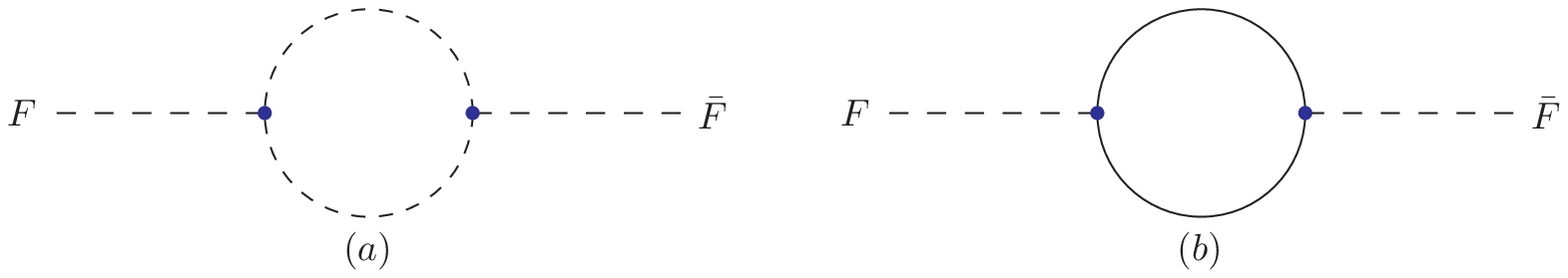,width=13.7cm}
\vspace*{-0.5cm}
\caption{Feynman diagrams contributing to the $F-$type soft terms.}
\label{fig3}
\end{figure}
\begin{figure}[htb]
\epsfig{file=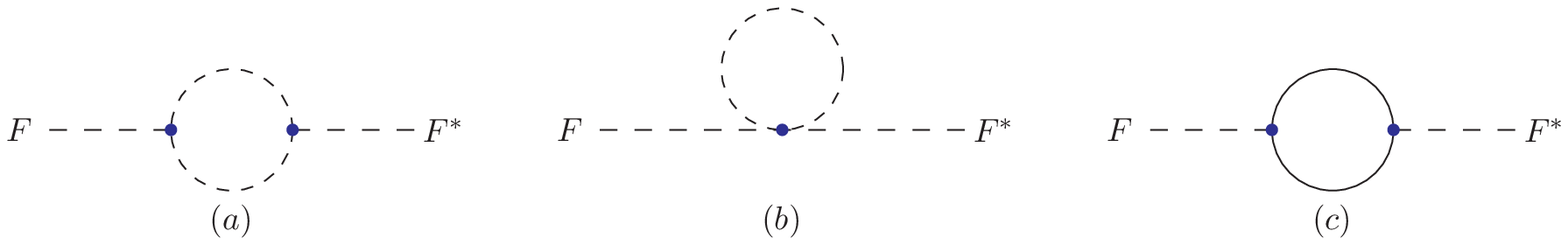,width=13.7cm}
\vspace*{-0.5cm}
\caption{Feynman diagrams contributing to the $D-$type soft terms.}
\label{F4}
\end{figure}
The messenger fields $(F,\bar F)$ get SUSY-breaking soft mass
terms through one-loop diagrams shown in Fig.~\ref{fig3}
and Fig.~\ref{F4}, which are given by
\begin{align}
m^2_{F\bar F}=&m^2_{F^*\bar F^*}=i(\Sigma_1+\Sigma_2+\Sigma_3)~, \\
m^2_{FF^*}=&m^2_{\bar F\bar F^*}=i(\Sigma_4+\Sigma_5+\Sigma_6)~,
\end{align}
where the $\Sigma$ functions are given by
\begin{align}
\Sigma_1=&\sum_{i}M_f\left[
\bar\lambda_a(\lambda_{ab}X+m_{ab})^*(A_{bi}A_{ic}^{\dagger})^*\lambda_c\frac{-i}{4\pi^2}f(m_{1i}^2,
M_f^2)+(A\rightarrow B, m_1\rightarrow m_2)
 \right. \nonumber \\
 &\left.
+\bar\lambda_a(C_{ai}C_{ib}^{\dagger})^*(\lambda_{bc}X+m_{bc})^*\lambda_c\frac{-i}{4\pi^2}f(m_{1i}^2,
M_f^2)+(C\rightarrow D, m_1\rightarrow m_2)\right],
\label{EqnA}\\
\Sigma_2=&\sum_iM_F\left[\bar\lambda_a(\lambda_{ab}X+m_{ab})^*(A_{bi}C_{ic}^{\dagger})^*\bar\lambda_c^*
\frac{-i}{4\pi^2}f(m_{1i}^2,
M_f^2) +
 \right. \nonumber \\
 &\left.
\lambda_a^*(A_{ai}C_{ib}^{\dagger})^*(\lambda_{bc}X+m_{bc})^*\lambda_c\frac{-i}{4\pi^2}f(m_{1i}^2,
M_f^2)
 +(A\rightarrow B, C\rightarrow D, m_1\rightarrow m_2)\right],
\label{EqnB}
\end{align}
Repeated indices should be summed over. Those two terms come from the
scalar loop (the left plot in Fig.~\ref{fig3}), and $\Sigma_1\propto
M_f$ is orders smaller than the second term. That is to say, the
holomorphic  corrections are dominant by the term proportional to the
low energy mass of itself.  The fermion loop contributes a terms
\begin{align}
\Sigma_3=&\sum_i[\lambda_aU_{Rai}U_{Lib}\bar\lambda_b\frac{i}{4\pi^2}M_{2i}M_ff(M_{2i},
M_f)]~.
\label{EqnC}
\end{align}
The non-holomorphic terms from Fig.~4 are given by
\begin{align}
\Sigma_4=&\sum_i[\bar\lambda_a(\lambda_{ab}X+m_{ab})(A_{bi}A_{ic}^{\dagger})^*(\lambda_{cd}X+m_{cd})^*\bar
\lambda_d^*\frac{-i}{4\pi^2}[\log m_{1i}^2-1] +(A\rightarrow B,
m_1\rightarrow m_2)],\cr
\Sigma_5=&\sum_i[\bar\lambda_a^*C_{ai}C_{ib}^{\dagger}\bar\lambda_b\frac{-i}{4\pi^2}m_{1i}^2\log
m_{1i}^2+(C\rightarrow D, m_1\rightarrow m_2)], \cr
\Sigma_6=&\sum_i[\bar\lambda_a^*U_{Lai}^{\dagger}U_{Lib}\bar\lambda_b\frac{i}{4\pi^2}M_{2i}^2(2\log
M_{2i}^2-1)].
\end{align}
The function $f(x,y)$ is defined as
\begin{align}
f(x, y)=\frac{x\log x-y\log y}{x-y}-1.
\end{align}
The messenger fields $(f,\bar f)$ have similar expressions except for a
flavor index.

\subsection{Soft Mass Terms from the Two-Loop Yukawa Mediation}

The two-loop corrections are dominant when $M=0$ in the hidden sector.
Now we calculate such two-loop corrections explicitly via
the wave function renormalization method.
We use the notation in Ref.~\cite{Delgado:2007rz}.
The superpotential is given by
\begin{align}
W=\lambda_{a}X\bar\phi_{ai}\phi_{ai}+\lambda_{a}\phi_{ai}\bar F_if
+\bar\lambda_{a}\bar\phi_{ai} F_i\bar f+M_FF_i\bar F_i+M_ff\bar f.
\end{align}
In the following calculations only one pair of secondary messenger
fields is
introduced, thus, $\ld_a=\ld=\bar \ld_a$. According to this method, we
have to calculate the threshold effect after integrating out the
heavy messengers. The soft terms can be extracted from the light
fields' wave-functions that get renormalziation from messengers,
which are given by
\begin{align}
m^2_{FF^*}=&-Z_F^{''}\frac{F^2}{M_h^2}, \cr
Z_F^{''}|_{Q=M_h}=&\frac{1}{4}\left[\beta_{\lambda}^{(+)}
 \frac{\partial(\Delta\gamma_{F})}
{\partial\lambda^2}-\Delta\beta_{\lambda}\frac{\partial\gamma_{F}^{(-)}
}{\partial\lambda^2}\right]_{Q=M_h}.
\end{align}
The beta-function, anomalous dimension and their change when
crossing the messenger scale are given by
\begin{align}
\Delta\gamma_{F}=&\frac{-2}{16\pi^2}\lambda^2, \quad
\gamma_{F}^{(-)}=0, \\
\beta_{\lambda}^{(+)}=&\frac{2\lambda^4}{16\pi^2}(N+2).
\end{align}
With this quantity we obtain
\begin{align}
m^2_{FF^*}=\frac{\lambda^4}{(16\pi^2)^2}(N+2)\frac{F^2}{M_h^2}.
\end{align}
Similarly, we have
\begin{align}
m^2_{ff^*}=\frac{\lambda^4}{(16\pi^2)^2}N(N+2)\frac{F^2}{M_h^2}.
\end{align}

\end{document}